\documentclass[iop]{emulateapj}
\usepackage{txfonts}
\usepackage{graphicx}
\usepackage{subfigure}
\usepackage{multirow}
\usepackage{verbatim}
\usepackage{longtable}
\usepackage{url}
\usepackage{float}
\usepackage{dcolumn}
\usepackage{hyperref}

\newcommand{\myemail}{carlziegler@unc.edu}

\def\lsim{\mathrel{\rlap{\lower4pt\hbox{\hskip1pt$\sim$}}
	\raise1pt\hbox{$<$}}}            	
\def\gsim{\mathrel{\rlap{\lower4pt\hbox{\hskip1pt$\sim$}}
	\raise1pt\hbox{$>$}}}            	
\shorttitle{Cool Subdwarf Multiplicity}
\shortauthors{Ziegler et al.}
\slugcomment{Submitted to ApJ}

\def\SB9{$S\!_{B^9}$}

\begin{document}

\title{Multiplicity of the Galactic Senior Citizens: \\ A high-resolution search for cool subdwarf companions}

\author{Carl Ziegler\altaffilmark{1}, Nicholas M. Law\altaffilmark{1}, Christoph Baranec\altaffilmark{2}, Reed L. Riddle\altaffilmark{3}, and Joshua T. Fuchs\altaffilmark{1}}

\altaffiltext{1}{Department of Physics and Astronomy, University of North Carolina at Chapel Hill, Chapel Hill, NC 27599-3255, USA}
\altaffiltext{2}{Institute for Astronomy, University of Hawai$\textquoteleft$i at M\={a}noa, HI 96720-2700, USA}
\altaffiltext{3}{Division of Physics, Mathematics, and Astronomy, California Institute of Technology, Pasadena, CA, 91125, USA}
\email{\myemail}

\begin{abstract}
Cool subdwarfs are the oldest members of the low-mass stellar population.  Mostly present in the galactic halo, subdwarfs are characterized by their low-metallicity.  Measuring their binary fraction and comparing it to solar-metallicity stars could give key insights into the star formation process early in the Milky Way's history.  However, because of their low luminosity and relative rarity in the solar neighborhood, binarity surveys of cool subdwarfs have suffered from small sample sizes and incompleteness.  Previous surveys have suggested that the binary fraction of red subdwarfs is much lower than for their main-sequence cousins.  Using the highly efficient Robo-AO system, we present the largest yet high-resolution survey of subdwarfs, sensitive to angular separations ($\rho$ $\ge$ 0$\farcs$15) and contrast ratios ($\Delta m_i \le$ 6) invisible in past surveys.  Of 344 target cool subdwarfs, 40 are in multiple systems, 16 newly discovered, for a binary fraction of $11.6\%\pm1.8\%$.  We also discovered 6 triple star systems for a triplet fraction of 1.7$\%\pm$0.7$\%$.  Comparisons to similar surveys of solar-metallicity dwarf stars gives a $\sim$3$\sigma$ disparity in luminosity between companion stars, with subdwarfs displaying a shortage of low-contrast companions.  We also observe a lack of close subdwarf companions in comparison to similar-mass dwarf multiple systems. 
\end{abstract}

\maketitle

\section{Introduction}
\label{Sec:Introduction}

Cool subdwarfs are the oldest members of the low-mass stellar population, with spectral types of K and M, masses between $\sim$0.6 and $\sim$0.08 M$_\sun$, and surface effective temperatures between ~4000 and ~2300 K \citep{kal09}.  First coined by \citet{kui39}, subdwarfs are the low-luminosity, metal-poor ([Fe/H] $<$ -1) spectral counterparts to the main sequence dwarfs. On a color-magnitude diagram, subdwarfs lie between white dwarfs and the main sequence \citep{adam15}. With decreased metal opacity, subdwarfs have smaller stellar radii and are bluer at a given luminosity than their main sequence counterparts \citep{sand59}.  These low-mass stars are members of the Galactic halo and have higher systematic velocities and proper motions than disk dwarf stars.  Traditionally subdwarfs have been identified using high proper motion surveys.  Although 99.7\% of stars in the galaxy are disk main sequence, statistically there are more subdwarfs in these high PM surveys \citep{reid05}.
	
The search for companions to stars of different masses gives clues to the star formation process, as any successful model must account for both the frequency of the multiple star systems and the properties of the systems.  In addition, monitoring the orbital characteristics of multiple star systems yields information otherwise unattainable for single stars, such as relative brightness and masses of the components \citep{good07a}, lending further constraints to mass-luminosity relationships \citep{chabrier00}

Old population II stars are important probes for the early history of star formation in the galaxy \citep{zhang13}.  The formation process of low mass stars remains less well understood than for solar-like stars.  Although multiple indications suggest they form as the low-mass tail of regular star formation \citep{bourke06}, other mechanisms have been proposed for some or all of these objects \citep{good07b, thies07, basu12}.  A firm binary fraction for low-metallicity cool stars could assist in constraining various formation models.  This again motivates the need for a comprehensive binarity survey, sensitive to small angular separations.

The multiplicity of main sequence dwarfs has been well explored in the literature.  A consistent trend that has purveyed is that the percentage of stars with stellar companions seems to depend on the mass of the stars.  For AB-type stars, \citet{pete12} used a sample of 148 stars to determine a companion fraction of $\sim$70\%.   For solar type stars (FGK-type), around 57\% have companions  \citep{duq91}, although \citet{rag10} have revised the fraction down to $\sim$46\%.  Fischer and Marcy (1992) looked at M-dwarfs and found a multiplicity fraction of 42$\pm$9\%.  More recently, \citet{janson12} find a binary fraction for late K- to mid M-type dwarfs of 27 $\pm$ 3$\%$ from a sample of 701 stars.  For late M-dwarfs, a slightly lower fraction was found by \citet{law06b} of 7$\pm$ 3$\%$.  Extending their previous study for mid/late M-type dwarfs, M5-M8, \citet{janson14} find a multiplicity fraction of 21\%-27\% using a sample of 205 stars.
	
While the multiplicity of dwarf stars has been heavily studied with comprehensive surveys, detailed multiplicity studies of low-mass subdwarfs have, historically, been hindered by their low luminosities and relative rarity in the solar neighborhood. Within 10 pc, there are three low-mass subdwarfs, compared to 243 main sequence stars \citep{monteiro06}.  Subsequently, multiplicity surveys of cool subdwarfs have been relatively small. The largest, a low-limit angular resolution search by \citet{zhang13} mined the Sloan Digital Sky Survey \citep{york00} to find 1826 cool subdwarfs, picking out subdwarfs by their PMs and identifying spectral type by fitting an absolute magnitude-spectral type relationship.  They find 45 subdwarfs multiple systems in total, with 30 being wide companions and 15 partially resolved companions.  When adjusting for the incompleteness of their survey, an estimate of the binary fraction of $>$10\% is predicted.  The authors note the need for a high spatial resolution imaging survey to search for close binaries ($<$100 AU) and put tighter constraints on the binary fraction of cool subdwarfs.

The high-resolution subdwarf surveys completed thus far have been comparatively small.  \citet{gizis00} detected no companions in a sample of eleven cool subdwarfs.  \citet{riaz08} similarly found no companions in a sample of nineteen M subdwarfs using the \textit{Hubble Space Telescope}.  \citet{lodieu09} reported one companion in a sample of 33 M type subdwarfs.  \citet{jao09} found four companions in a sample of 62 cool subdwarf systems.  With the high variance in small number statistics, the relationship between dwarf and subdwarf multiplicity fractions remains inconclusive.

We present here the largest high resolution cool subdwarf multiplicity survey yet performed, making use of the efficient Robo-AO system.   The Robo-AO system allows us to detect more cool and close companion stars in a much larger sample size than previously possible. This survey combines previously known wide proper-motion pairs, spectroscopic binaries, and high angular resolution images able to detect companions with $\rho$ $\ge$ 0$\farcs$15 and $\Delta m_i \le$ 6.

The paper is organized as follows.  In Section \ref{sec:Targetselection} we describe the target selection, the Robo-AO system, and follow-up observations. In Section \ref{sec:Data} we describe the Robo-AO data reduction and the companion detection and analysis.  In Section \ref{sec:Discoveries} we describe the results of this survey, including discovered companions, and compare to similar dwarf surveys.  The results are discussed in Section \ref{sec:Discussion} and put in context of previous literature.  We conclude in Section \ref{sec:Conclusions}.
	
\section{Survey Targets and Observations}
\label{sec:Targetselection}
\subsection{Sample Selection}

\begin{figure}
\centering
\includegraphics[width=245pt]{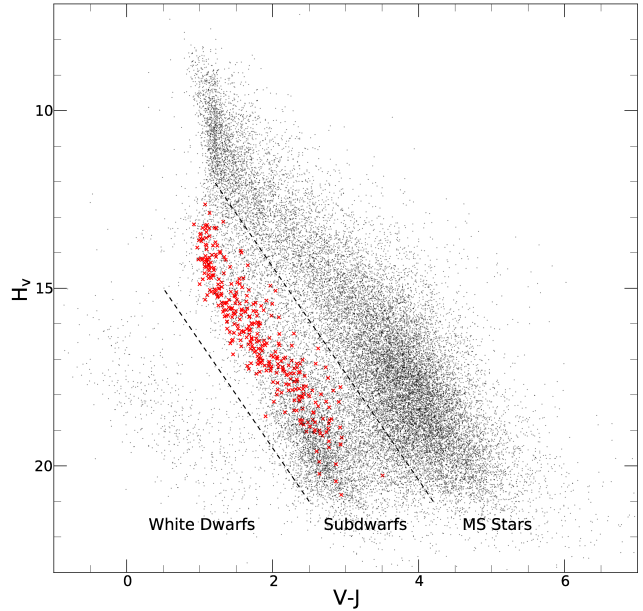}
\caption{Reduced proper motion diagram of the complete rNLTT \citep{gould03}, with our observed subdwarfs in red \textit{X}'s.  The discriminator lines between solar-metallicity dwarfs, metal-poor subdwarfs, and white dwarfs are at $\eta$ = 0 and 5.15, respectively, and with \textit{b}=$\pm$30.  The subdwarfs plotted make use of the improved photometry of \citet{marshall07}.}
\label{fig:redprop}
\end{figure}

\begin{figure}
\centering
\includegraphics[width=241pt]{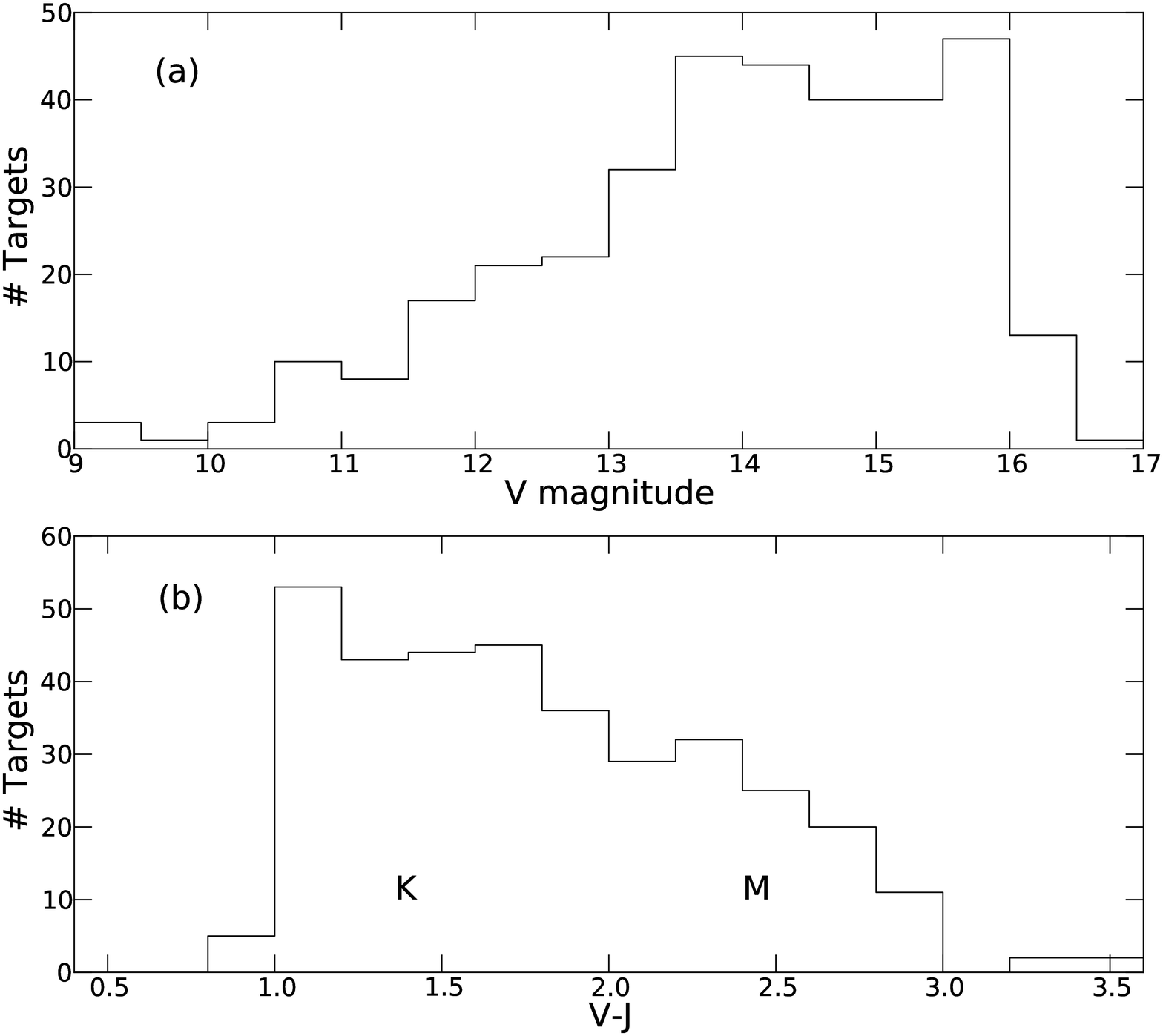}
\caption{(\textit{a}) Histogram of magnitudes in V band of the 348 observed subdwarfs. (\textit{b}) Histogram of the $(V - J)$ colors of the observed subdwarf sample, with approximate spectral types regions K and M marked.  Both plots use the photometry of \citet{marshall07}}
\label{fig:hist_colorvmag}
\end{figure}

\begin{table}
\renewcommand{\arraystretch}{1.3}
\begin{longtable}{ll}
\caption{\label{tab:survey_specs}The specifications of the Robo-AO subdwarf survey}
\\
\hline
Filter & Sloan \textit{i}\textsuperscript{$\prime$}-band \\
FWHM resolution   	& 0$\farcs$15 \\
Field size & 44\arcsec $\times$ 44\arcsec\\
Detector format       	& 1024$^2$ pixels\\
Pixel scale & 43.1 mas / pix\\
Exposure time & 120 seconds \\
Subdwarf targets    	& 344 \\
Targets observed / hour & 20\\
Observation dates & September 1 2012 --\\ &  August 21 2013\\
\hline
\label{tab:specs}
\end{longtable}
\end{table}

\begin{figure*}[!htb]
\centering
\includegraphics*[width=400pt]{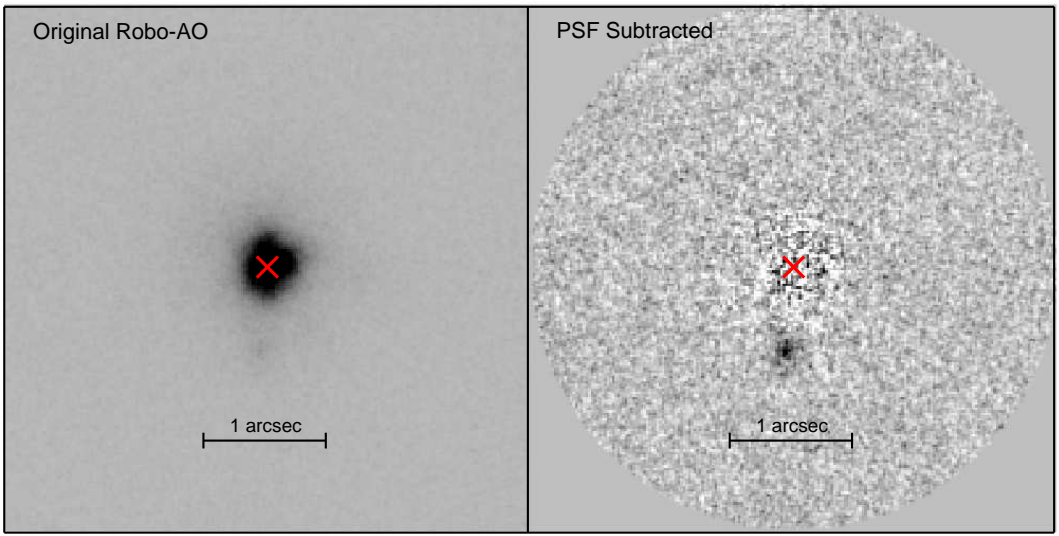}
\caption{Example of PSF subtraction on NLTT31240 with companion separation of 0$\farcs$74.  The red X marks the position of the primary star's PSF peak. Successful removal of PSF leaves residuals consistent with photon noise.}
\label{fig:psf}
\end{figure*}

We selected targets from the 564 spectral type F- through M-subdwarfs studied by Marshall (2007). These targets were selected from the New Luyten Two-Tenths (NLTT) catalog \citep{luyten79, luyten80} of high proper motion stars ($>$0.18 arcsec/year) using a reduced proper motion diagram (RPM).  To distinguish subdwarf stars from their solar-metallicity companions on the main sequence, the RPM used a $(V-J)$ optical-infrared baseline, a technique first used by \citet{salim02}, rather than the shorter $(B-R)$ baseline used by Luyten.  This method uses the high proper motion as a proxy for distance and the blueness of subdwarfs relative to equal luminosity dwarf stars to separate out main sequence members of the local disk and the halo subdwarfs \citep{marshall08}.  The reduced proper motion, H$_M$, is defined as 
\begin{equation}
H_{M} = m + 5log\mu + 5
\end{equation}
where m is the apparent magnitude and $\mu$ is the proper motion in $\arcsec$/yr.  The discriminator, $\eta$, developed by Salim \& Gould to separate luminosity classes, is defined as
\begin{equation}
\eta(H_{V},V-J,\sin b)=H_{V}-3.1(V-J) - 1.47|\sin b| - 7.73
\end{equation}
where \textit{b} is the Galactic latitude.  The reduced proper motion diagram for the revised NLTT (rNLTT) catalog \citep{gould03} and our subdwarf targets is presented in Figure$~\ref{fig:redprop}$.  The improved photometry of \citet{marshall07} placed 12 of the original suspected subdwarfs outside the subdwarf sequence.  These stars were rejected from our sample.  Of the 552 subdwarfs confirmed by Marshall, a randomly-selected sample of 348 K- and M-subdwarfs were observed by Robo-AO when available between other high priority surveys.  The V-band magnitudes and $(V-J)$ colors of the observed subdwarf sample are shown in Figure$~\ref{fig:hist_colorvmag}$.

\subsection{Observations}
\subsubsection{Robo-AO}
We obtained high-angular-resolution images of the 348 subdwarfs during 32 separate nights of observations between 2012 September 3 and 2013 August 21 (UT).  The observations were performed using the Robo-AO laser adaptive optics system \citep{baranec13, baranec14, riddle12} mounted on the Palomar 60 inch telescope.  The first robotic laser guide star adaptive optics system, the automatic Robo-AO system can efficiently observe large, high-resolution surveys. All images were taken using the Sloan \textit{i}\textsuperscript{$\prime$}-band filter \citep{york00} and with exposure times of 120 s.  Typical seeing at the Palomar Observatory is between 0$\farcs$8 and 1$\farcs$8, with median around 1$\farcs$1 \citep{baranec14}.  The typical FWHM (diffraction limited) resolution of the Robo-AO system is 0$\farcs$12-0$\farcs$15. Specifications of the Robo-AO system are summarized in Table$~\ref{tab:specs}$.

The images were reduced by the Robo-AO imaging pipeline described in \citet{law06a, law06b, law09, law14}. The EMCCD frames are dark-subtracted and flat-fielded and then, using the Drizzle algorithm \citep{fruchter02}, stacked and aligned, while correcting for image motion using a bright star in the field. The algorithm also introduces a factor-of-two up-sampling to the images. Since the subdwarf targets are in relatively sparse stellar fields, for the majority of the images the only star visible is the target star and was thus used to correct for the image motion.

\subsubsection{Keck LGS-AO}
Six candidate multiple systems were selected for re-imaging by the NIRC2 camera behind the Keck II laser guide star adaptive optics system \citep{wiz06, vandam06}, located in Maunakea, Hawaii, on 2014 August 17 (UT) to confirm possible companions.  The targets were selected for their low significance of detectability, either because of low contrast ratio or small angular separation.  The observations were done in the K\textsuperscript{$\prime$} and H bands with three 90 s exposures for two targets and three 30 s for five targets in a 3-position dither pattern that avoided the noisy, lower-left quadrant.  We used the narrow camera setting (0$\farcs$0099/px), which gave a single-frame field of view of 10\arcsec $\times$ 10\arcsec.

\subsubsection{SOAR Goodman Spectroscopy}
We took spectra of 24 of the subdwarfs using the Southern Astrophysical Research Telescope and the Goodman Spectrograph \citep{clemens04} on 2014 July 15. We observed twelve targets with companions and twelve single stars as reference.  The spectra were taken using a 930 lines/mm grating with 0.42 \AA/pixel, a 1$\farcs$07 slit, and exposure times of 480 s.

\section{Data Reduction and Analysis}
\label{sec:Data}

\subsection{Robo-AO Imaging}
\subsubsection{Target Verification}

To verify that each star viewed in the image is the desired subdwarf target, we created Digital Sky Survey cutouts of similar angular size around the target coordinates.  Each image was then manually checked to assure no ambiguity in the target star.  The vast majority of the targets are in relatively sparse stellar regions.  Four of the target stars in crowded fields whose identification was ambiguous were discarded, leaving 344 verified subdwarf targets.

\subsubsection{PSF Subtraction}

To locate close companions, a custom locally optimized point spread function (PSF) subtraction routine \citep{law14} based on the Locally Optimized combination of Images algorithm \citep{lafreniere07} was applied to centered cutouts of all stars.  Other subdwarf observations taken at similar times were used as references, as it is unlikely to have a companion found in the same position for two different targets.  For each target image and for 20 reference images selected as the closest to the target image in observation time, the region around the star was subdivided into polar sections, five up-sampled pixels in radius and 45$^{\circ}$ in angle.  A locally optimized estimate of the PSF for each section was then generated using a linear combination of the reference PSFs.  The algorithm begins with an average over the reference PSFs, then uses a downhill simplex algorithm to optimize the contributions from each reference image to find the best fit to the target image.  The optimization is done on several coincident sections simultaneously to minimize the probability of subtracting out a real companion, with only the central region outputted to the final PSF.  This also provides smoother transitions between adjacent sections as many of the image pixels were shared in the optimization.

After iterating over all sections of the image, the final PSF is an optimal local combination of all the reference PSFs.  This final PSF is then subtracted from the original reference image, leaving residuals that are consistent with photon noise. Figure$~\ref{fig:psf}$ shows an example of the PSF subtraction performance.

We manually checked the final subtracted images for close companions detections ($>$5$\sigma$).  The initial search was limited to a detection radius of 1$\arcsec$ from the target star. We subsequently performed a secondary search out to a radius of 2$\arcsec$.

\subsubsection{Imaging Performance Metrics}
\label{sec:imageperf}

The two dominant factors that effect the image performance of the Robo-AO system are seeing and target brightness.  To further classify the image performance for each target an automated routine was ran on all images.  Described in detail in \citet{law14}, the code uses two Moffat functions fit to the PSF to separate the widths of the core and halo.  We found that the core size was an excellent predictor of the contrast performance, and used it to group targets into three levels (low, medium and high).  Counter-intuitively, the PSF core size decreases as image quality decreases.  This is caused by poor S/N on the shift-and-add image alignment used by the EMCCD detector.  The frame alignment subsequently locks onto photo noise spikes, leading to single-pixel-sized spikes in the images \citep{law06b,law09}. The images with diffraction limited core size ($\sim$0.15\arcsec) were assigned to the high-performance group, with smaller cores assigned to lower-performance groups.

Using a companion-detection simulation with a group of representative targets, we determine the angular separation and contrast consistent with a 5$\sigma$ detection.  For clarity, the contrast curves of the simulated targets are fitted with functions of the form $a - b/(r - c)$ (where \textit{r} is the radius from the target star and \textit{a, b,} and \textit{c} are fitting variables).  Contrast curves for the three performance groups are shown in Section$~\ref{sec:Discoveries}$.

\begin{figure}
\centering
\includegraphics[width=250pt]{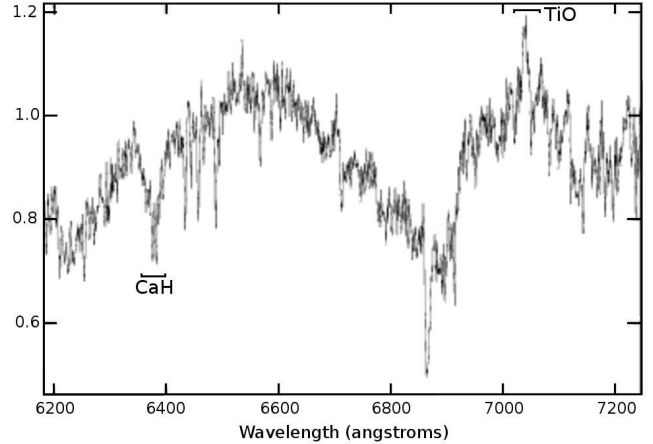}
\caption{The extracted spectra for NLTT52532 showing subdwarf characteristics,  most apparent the weakness of the 7050\AA TiO band and strength of the 6380\AA CaH band.  The y-axis is given in normalized arbitrary flux units.}
\label{fig:spect}
\end{figure}

\begin{table}
\begin{center}
\caption{Full SOAR Spectroscopic Observation List}
\begin{tabular}{cccc}
\hline
\hline
\noalign{\vskip 3pt}  
\text{NLTT} & \text{m$_v$} & \text{ObsID} & \text{Companion?}\\ [0.2ex]
\hline \\ [-1.5ex]
2205 & 14.0 & 2014 Jul 14 & yes\\
7301 & 14.9 & 2014 Jul 14 & yes\\
7914 & 14.3 & 2014 Jul 14 & yes\\
9597 & 12.0 & 2014 Jul 14 & \\
9898 & 14.2 & 2014 Jul 14 & \\
10022 & 15.8 & 2014 Jul 14 & \\
10135 & 15.7 & 2014 Jul 14 & \\
33971 & 12.8 & 2014 Jul 14 & \\
37342 & 14.4 & 2014 Jul 14 & yes\\
37807 & 12.0 & 2014 Jul 14 & \\
40022 & 13.9 & 2014 Jul 14 & \\
40313 & 13.7 & 2014 Jul 14 & \\
41111 & 13.7 & 2014 Jul 14 & \\
44039 & 11.5 & 2014 Jul 14 & \\
44568 & 12.3 & 2014 Jul 14 & \\
49486 & 16.0 & 2014 Jul 14 & yes\\
50869 & 15.8 & 2014 Jul 14 & \\
52377 & 14.5 & 2014 Jul 14 & yes\\
52532 & 15.5 & 2014 Jul 14 & yes\\
53255 & 15.0 & 2014 Jul 14 & yes\\
55603 & 12.1 & 2014 Jul 14 & yes\\
56818 & 14.0 & 2014 Jul 14 & yes\\
57038 & 13.9 & 2014 Jul 14 & yes\\
58812 & 14.9 & 2014 Jul 14 & yes\\
\end{tabular}
\label{tab:soar}
\end{center}
\end{table}

\subsubsection{Contrast Ratios}

For wide companions, the binaries' contrast ratio was determined using aperture photometry on the original images. The aperture size was determined uniquely for each system based on separation and the presence of non-associated background stars.

For close companions, the estimated PSF was used to remove the blended contributions of each star before aperture photometry was performed.  The locally optimized PSF subtraction algorithm attempts to remove the flux from companions using other reference PSFs with excess brightness in those areas.  For detection purposes, we use many PSF core sizes for optimization, and the algorithm's ability to remove the companion light is reduced. However, the companion is artificially faint as some flux has still been subtracted.  To avoid this, the PSF fit was redone excluding a six-pixel-diameter region around the detected companion.  The large PSF regions allow the excess light from the primary star to be removed, while not reducing the brightness of the companion.

\subsubsection{Separation and Position Angles}

Separation angles were determined from the raw pixel positions.  Uncertainties were found using estimated systematic errors due to blending between components.  Typical uncertainty in the position for each star was 1-2 pixels.  Position angles were calculated using a distortion solution produced using Robo-AO measurements for a globular cluster.\footnote{S. Hildebrandt (2013, private communication).}

\subsection{Previously Detected Binaries}

To further realize our goal of a comprehensive cool subdwarf survey, we included in our statistics previously confirmed binary systems in the literature with separations outside of our field of view.  Common proper motion is a useful indicator of wider binary systems.  Wide ($>$30\arcsec) common proper motion companions among our target subdwarfs were previously identified in the Revised New Luyten Two-Tenths (rNLTT) catalog \citep{salim02,chan04}, and a search by \citet{lopez12} of the Lepine and Shara Proper Motion-North catalog \citep{lepine05}. 

The target list was also cross-checked against the {\it Ninth Catalogue of Spectroscopic Binary Orbits} \citep[\protect\SB9]{pourbaix04}, a catalogue of known spectroscopic binaries available online.\footnote{http://sb9.astro.ulb.ac.be/}  While these systems were included in the total subdwarf binary numbers, the compilatory nature of this catalogue leaves some uncertainty in the completeness of the spectroscopic search.

\begin{figure*}
\centering
\includegraphics[width=400pt]{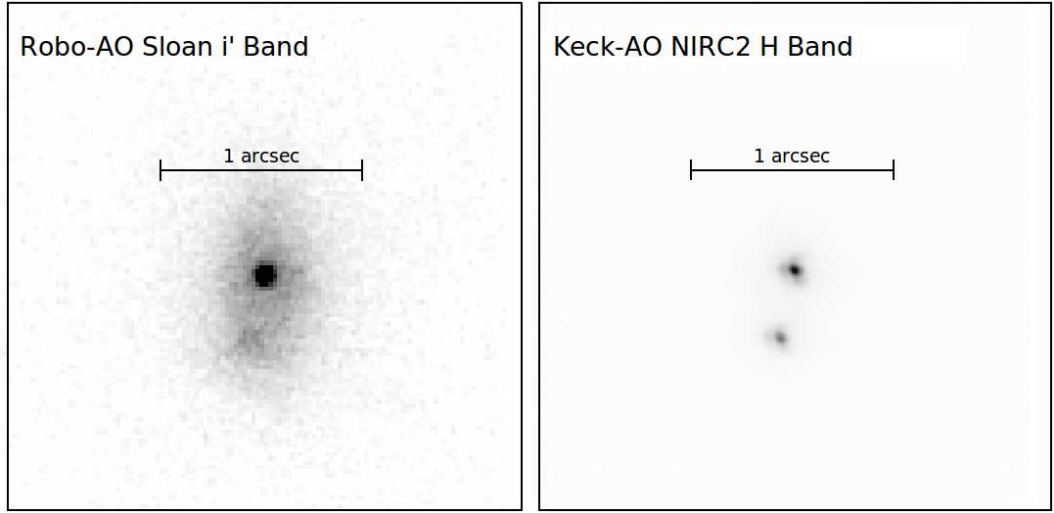}
\caption{Keck-AO confirming the Robo-AO companion to NLTT52532. Exposure time for the Robo-AO image is 120 s and for the Keck-AO image is 90 s.}
\label{fig:keck}
\end{figure*}

\subsection{Spectroscopy}

To further verify that the targets selected are cool subdwarfs, we took spectra of 7$\%$ of the total survey and 31$\%$ of the candidate companion systems.  Past spectroscopic studies of cool subdwarfs at high resolution have proven difficult as, at the low temperatures present, a forest of molecular absorption lines conceals most atomic lines used in spectral analysis.  Subdwarfs can be classified spectroscopically using two molecular lines \citep{gizis97}. Comparing titanium oxide (TiO) bands to metal hydride bands (typically CaH in M subdwarfs), Gizis classified two groups, the intermediate and extreme subdwarfs.  As the metallicity decreases, the TiO adsorption also decreases, but the CaH remains largely unaffected for a given spectral type. This classification system was expanded and revised to include ultra subdwarfs by \citet{lepine07}, who introduced the new useful parameter $\zeta_{TiO/CaH}$.
	
Spectra were taken for wavelengths 5900-7400\AA, and reduced (dark-subtracted and flat-fielded) using IRAF reduction packages, particularly the onedspec.apall to extract the trace of the spectrum and onedspec.dispcor for applying the wavelength calibration. A Fe+Ar arc lamp was recorded for wavelength calibration. All observed target subdwarfs were confirmed to show the spectral characteristics of subdwarf stars described above, specifically the reduced band strength of 7050\AA TiO5.  An example of the extracted spectra is given in Figure$~\ref{fig:spect}$.  The full observation list for SOAR is given in Table$~\ref{tab:soar}$.

\begin{figure}
\centering
\includegraphics[width=245pt]{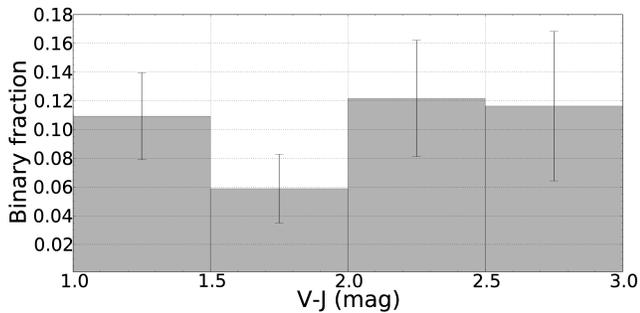}
\caption{Binary fraction of the target subdwarfs binned by their $(V-J)$ color.}
\label{fig:binary_frac_color}
\end{figure}

\begin{figure}
\centering
\includegraphics[width=245pt]{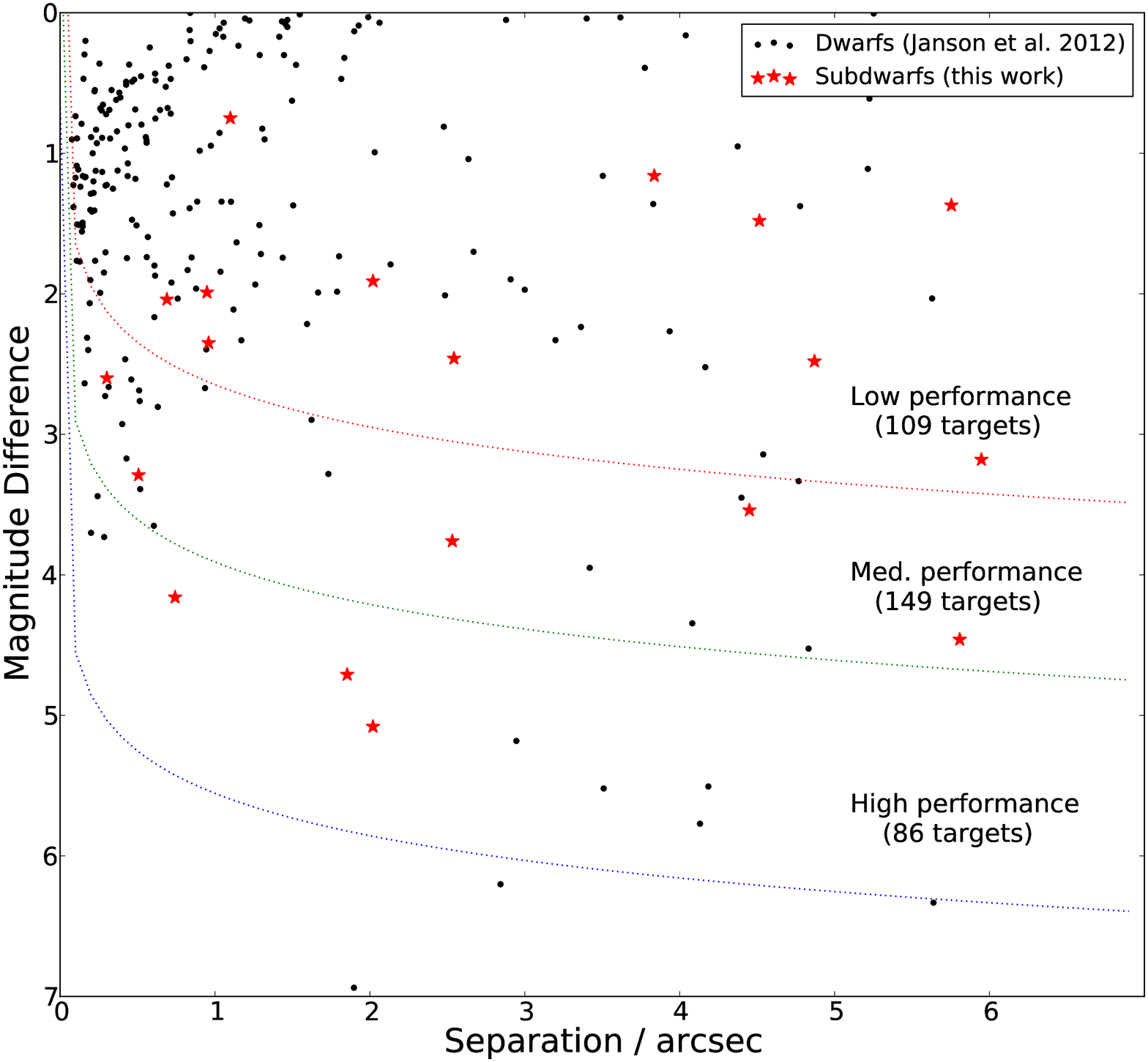}
\caption{Comparison of the separation and the magnitude difference in the i-band between our subdwarf companions $(<$6$\arcsec$) and the dwarf companions found by \citet{janson12}.  The detectable magnitude ratios for our image performance groups are also plotted, with the number of observed subdwarfs targets in each image performance group, as described in Section \ref{sec:imageperf}.}
\label{fig:dwarfscomp}
\end{figure}

\subsection{Candidate Companion Follow-ups}

With either high contrast ratio or small angular separation, six candidate subdwarf binary systems with low detection significance ($<$6$\sigma$) were selected for follow-up imaging using Keck II.  One low-probability candidate companion star was rejected after followups using Keck II, an apparent close ($\rho\simeq0.15\arcsec$) binary to NLTT50869, probably resulting from a cosmic ray on the original Robo-AO image.  A wider binary to NLTT50869, with high detection significance, was not in the image field of view.  Outside of the six target stars with low significant companions, another candidate companion star, NLTT4817, was observed and had no companion inside the field of view of the Keck II image, however had a high significant companion ($>$7$\sigma$) in the Robo-AO field of view.  An example of the Keck II images and the Robo-AO images is given in Figure$~\ref{fig:keck}$.  The full Keck II observations are listed in Table$~\ref{tab:keck}$, with the last column indicating the presence of the low detection significance companion.

\begin{table}
\renewcommand{\arraystretch}{1.3}
\begin{center}
\caption{Full Keck-AO Observation List}
\begin{tabular}{cccc}
\hline
\hline
\noalign{\vskip 1pt}  
\text{NLTT} & \text{m$_v$} & \text{ObsID} & \text{Low-sig. Companion?}\\ [0.2ex]
\hline\
4817 & 11.4 & 2014 Aug 17 & \\
7914 & 14.3 & 2014 Aug 17 & yes \\
50869 & 15.8 & 2014 Aug 17 &  \\
52377 & 14.5 & 2014 Aug 17 & yes \\
52532 & 15.5 & 2014 Aug 17 & yes \\
53255 & 15.0 & 2014 Aug 17 & yes \\
56818 & 14.0 & 2014 Aug 17 & yes \\
\end{tabular}
\label{tab:keck}
\end{center}
\end{table}

\begin{figure*}
\centering
\includegraphics[width=500pt]{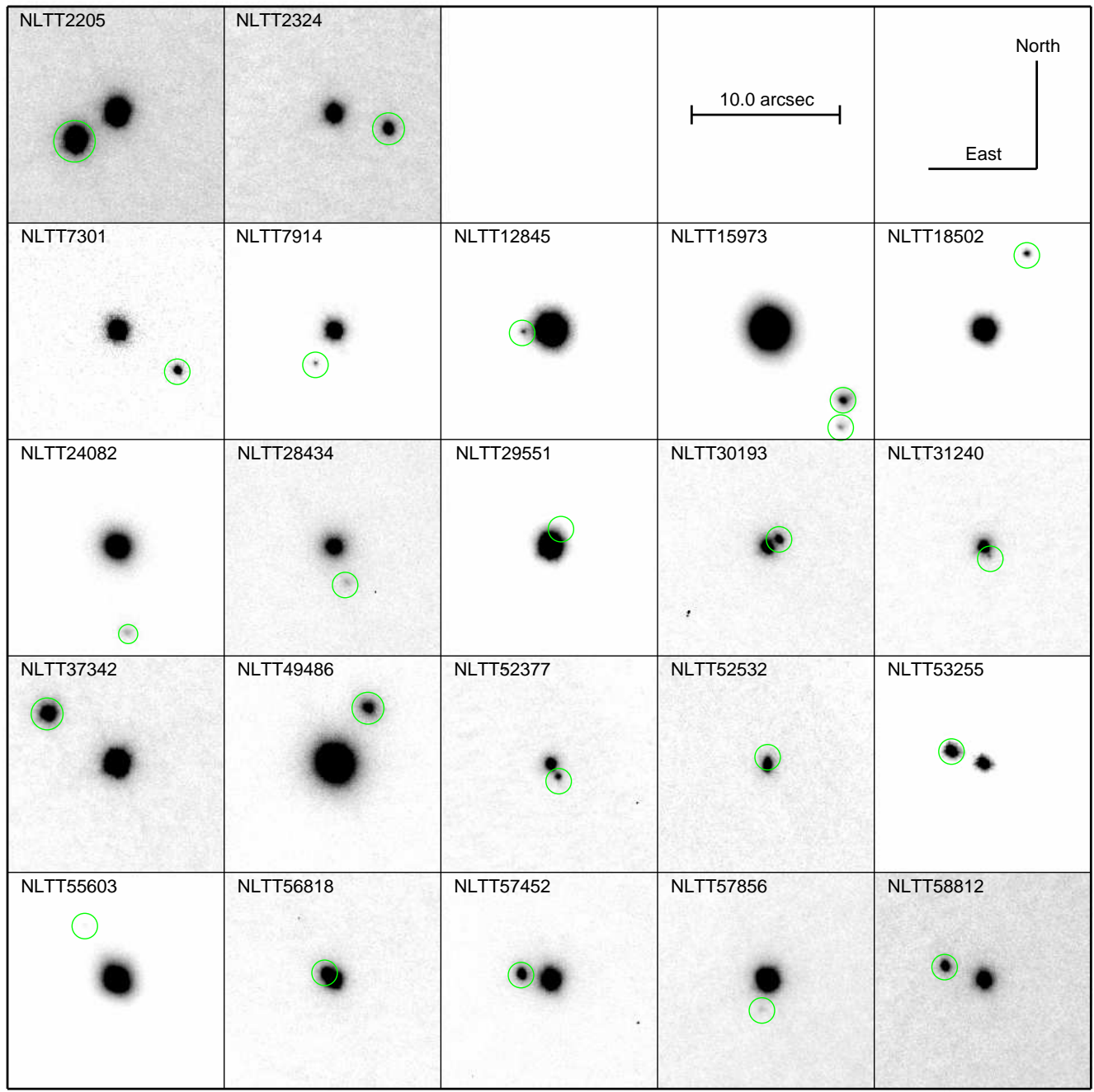}
\caption{Gray scale cutouts of the 22 multiple star systems with separations $<$7$\arcsec$ resolved with Robo-AO.  The angular scale and orientation is similar for each cutout.}
\label{fig:cutouts}
\end{figure*}

\begin{figure}
\centering
\includegraphics[width=245pt]{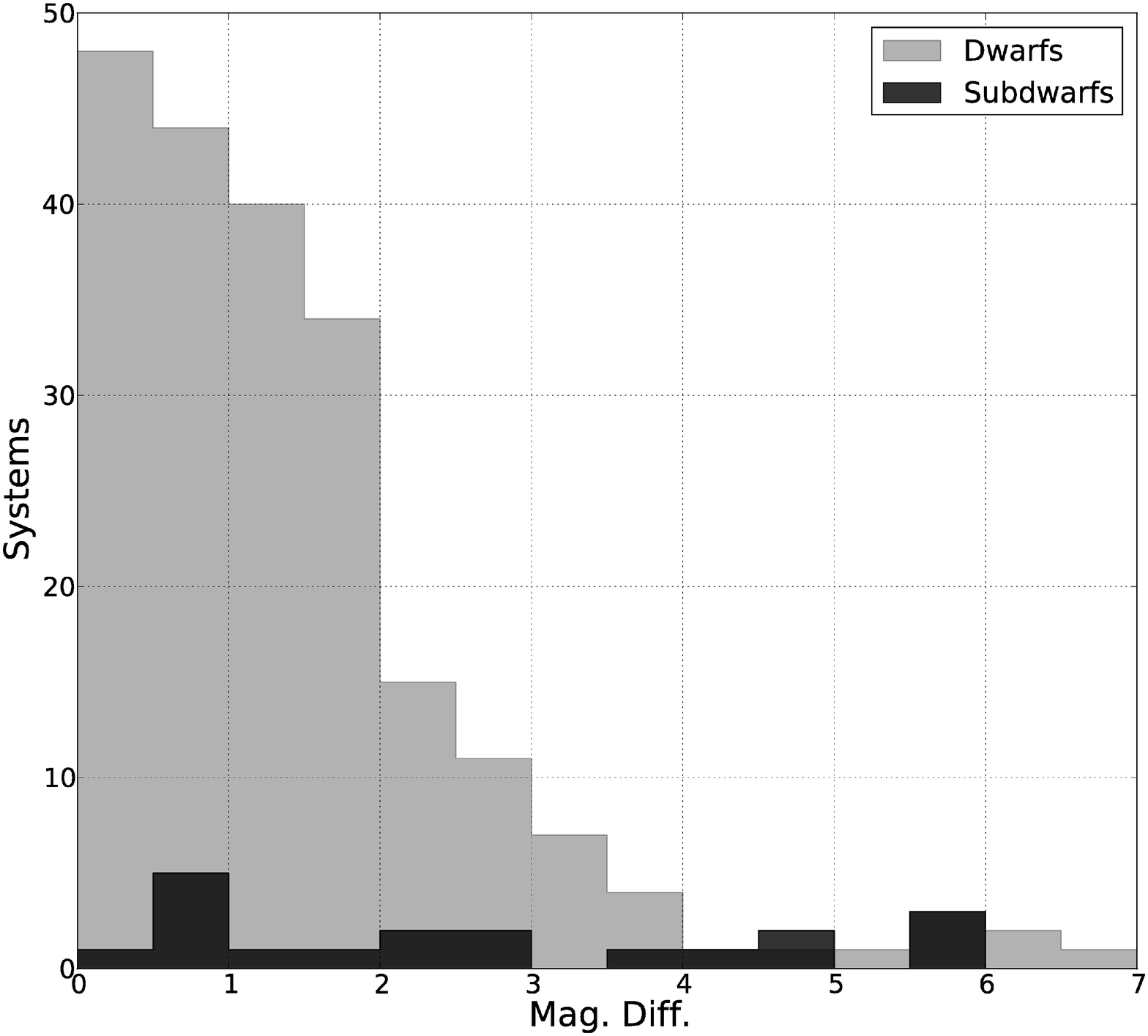}
\caption{Histogram of the magnitude difference in the i-band between all our subdwarf companions and the dwarf companions found by \citet{janson12}.}
\label{fig:hist_mag}
\end{figure}

\section{Discoveries}
\label{sec:Discoveries}

Of the 344 verified subdwarf targets observed, 40 appear to be in multiple star systems for an apparent binary fraction of 11.6\%$\pm$1.8$\%$, where the error is based on Poissonian statistics \citep{burgasser03}.  This count includes 6 multiple systems first recorded in the NLTT, 13 systems first recorded in the rNLTT, 1 wide binary found in the LSPM \citep{lopez12}, 6 spectroscopic binaries, and 16 newly discovered multiple systems.  We also found four new companions to already recorded binary systems, including two new triple systems, for a total of 6 triple star systems, for a triplet fraction of 1.7\%$\pm$.7$\%$.  One quarter (26$\%$) of the companions would only be observable in a high-resolution survey ($<$2.0\arcsec sep).  The binary fraction of the target stars binned by their $(V-J)$ color is given in Figure$~\ref{fig:binary_frac_color}$. Cutouts of the closest 22 multiple star systems are shown in Figure$~\ref{fig:cutouts}$. Measured companion properties are detailed in Table$~\ref{tab:measurements}$.

\subsection{Probability of Association}

The associations of all discovered and previously recorded companions were confirmed using the Digitized Sky Survey (DSS) \citep{reid91}.  Since all the targets have high proper motions, if not physically associated the systems would have highly apparent shifts in separation and position angle over the past two decades.  For the widely separated systems with both stars visible in the DSS, we checked the angular separation in the DSS and our survey to confirm relatively constant separation.  For closely separated systems where both stars are merged in the DSS, we looked for a background star at the DSS position that does not appear in our images.

In addition, since our stars appear in relatively sparse stellar regions in the sky, well outside the Galactic disk, the probability of a background star appearing in a close radius to our observed star is low.  Using the total number of known non-associated stars in our images, than we expect over all target stars in our survey 1.2 background stars within a radius of 2$\farcs$5 of any of our target stars, compared to 10 stars observed in that range.

\subsection{Photometric Parallaxes}

Very few subdwarfs in our sample have accurate parallax measurements.  Only 43 of the targets have published parallaxes, most with significant measurement errors.  To estimate the distances to our subdwarf targets, we employed an expression for M$_{R}$=$\fint (R-I)$ estimated by \citet{siegel02} using a color-magnitude diagram and the photometric measurements by \citet{marshall07}.
	
	The polynomial fit found by Siegel for subdwarfs with measured parallaxes and an estimated mean [Fe/H] of -1.2, and with the \citet{lutz73} correction, is
\begin{equation}
M_{R} = 2.03 + 10 \times (R-I) - 2.21 \times (R-I)^2
\end{equation}
The color-absolute magnitude relation has an uncertainty of $\sim$0.3 mag.  In all cases, the published parallax errors are much larger than photometric errors of $<$0.03 mag.  The estimated distances for the primary stars in the subdwarf multiple systems are listed in Table$~\ref{tab:measurements}$.

\section{Discussion}
\label{sec:Discussion}
\subsection{Comparison to Main-Sequence Dwarfs}

With comparable sample size and spectrum types, the cool dwarf survey of \citet{janson12} is a useful metal-rich analog to this work. The most striking disparity between the two samples is the lack of low-contrast ($\Delta m_i \leq$2), close ($\rho \leq 1 \arcsec$) companions to the subdwarf stars, a regime heavily populated by solar-metallicity dwarf companions.  This is clearly seen in a plot of the companion's magnitude difference versus angular separation for the two populations, as in Figure$~\ref{fig:dwarfscomp}$.  

The dissimilarity between contrast ratios between dwarfs and subdwarfs is further illustrated in Figure$~\ref{fig:hist_mag}$. A two sample Kolmogrov-Smirnov test rejects the null hypothesis that the two populations are similar at a confidence of $\sim$2.8$\sigma$.

\begin{table*}
\renewcommand{\arraystretch}{1.2}
\setlength{\tabcolsep}{6pt}
\caption{Multiple subdwarf systems resolved using Robo-AO and previously detected systems}
\small
\centering
\begin{tabular}{l c c c c c c c c c}
\hline
\hline
 NLTT &  Comp & \textit{$m_v$}\footnote{\citep{marshall07}} &  ObsID &  $\Delta$ \textit{i}\textsuperscript{$\prime$} &  $\rho$  &  $\rho$  &  P.A.  &  Dist&  Prev Det?\\
 & NLTT &  (mag) &  & (mag) &  ($\arcsec$) &  (AU) &  (deg.) &  (pc)\\
\hline
2045AB & \nodata & 13.5 & 2013 Aug 15 & \nodata & \nodata & \nodata & \nodata & 183.3$\pm$21.0 & SB9\\
2205AB & 2206 & 13.9 & 2013 Aug 15 & 0.18 & 3.37 & 475.5$\pm$54.3 & 123$\pm$2 & 140.9$\pm$16.1 & L79\\
2324AB & 2325 & 15.7 & 2013 Aug 16 & 1.16 & 3.84 & 138.8$\pm$15.9 & 254$\pm$2 & 36.1$\pm$4.1 & L79\\
2324AC & \nodata & 15.7 & 2013 Aug 16 & 4.14 & 23.48 & 847.8$\pm$96.2 & 159$\pm$2 & 36.1$\pm$4.1 & \\
4817AB & 4814 & 11.4 & 2012 Sep 3 & 4.30 & 24.59 & 3615$\pm$413 & 218$\pm$2 & 147$\pm$16.8 & S02\\
7301AB & 7300 & 14.9 & 2012 Sep 3 & 2.48 & 4.87 & 105.7$\pm$12.1 & 57$\pm$2 & 21.7$\pm$2.5 & S02\\
7914AB & \nodata & 14.3 &  2012 Sep 3 & 3.76 & 2.53 & 424.4$\pm$48.5 & 150$\pm$2 & 167.6$\pm$19.2 & \\
10536AB & 10548 & 11.2 & 2013 Aug 15 & \nodata & 185.7 & 30633$\pm$3501 & 85.5 & 164.9$\pm$18.9 & S02\\
11015AB & 11016 & 16.3 & 2013 Aug 16 & 0.94 & 9.24 & 1399$\pm$160 & 57$\pm$2 & 151.3$\pm$17.3 & S02\\
12845AB & \nodata & 10.6 & 2012 Oct 3 & 4.71 & 1.85 & 149.4$\pm$17.1 & 92$\pm$2 & 80.6$\pm$9.2 & \\
15973AB & 15974 & 9.3 &  2012 Oct 7 & 3.47 & 6.88 & 303.1$\pm$34.6 & 227$\pm$2 & 44$\pm$5.0 & S02\\
15973AC & \nodata & 9.3 & 2012 Oct 7 & 5.02 & 8.23 & 362.2$\pm$41.1 & 217$\pm$2 & 44$\pm$5.0 & \\
17485AB & \nodata & 11.9 & 2012 Oct 10 & \nodata & \nodata & \nodata & \nodata & 191.3$\pm$21.9 & SB9\\
18502AB & \nodata & 12.2 & 2013 Jan 19 & 3.18 & 5.95 & 1262$\pm$144 & 331$\pm$2 & 212.1$\pm$24.3 & \\
18798AB & 18799 & 14.5 & 2013 Jan 19 & 3.12 & 12.82 & 2270$\pm$259 & 172$\pm$2 & 177$\pm$20.2 & S02\\
19210AB & 19207 & 11.2 & 2013 Jan 20 &  & 102.5 & 18468$\pm$2110 & 285.4 & 180.2$\pm$20.6 & S02,SB9\\
20691AB & \nodata & 9.6 & 2013 Jan 19 & \nodata & \nodata & \nodata & \nodata & 70.6$\pm$8.1 & SB9\\
21370AB & \nodata & 13.7 & 2013 Jan 19 & 2.46 & 19.83 & 6603$\pm$755 & 322$\pm$2 & 332.9$\pm$38.1 & SB9\\
24082AB & \nodata & 13.1 & 2013 Jan 19 & 4.46 & 5.81 & 1683$\pm$192 & 187$\pm$2 & 289.7$\pm$33.1 & \\
24082AC & \nodata & 13.1 & 2013 Jan 19 & 4.17 & 12.00 & 3476$\pm$397 & 267$\pm$2 & 289.7$\pm$33.1 & \\
25234AB & 25233 & 13.2 &2013 Jan 18 & 3.05 & 8.29 & 1175$\pm$134 & 287$\pm$2 & 141.7$\pm$16.2 & S02\\
28434AB & \nodata & 14.9 & 2013 Jan 17 & 2.46 & 2.54 & 652.9$\pm$74.6 & 202$\pm$2 & 256.7$\pm$29.3 & \\
29551AB & \nodata & 11.5 & 2012 Sep 3 & 3.29 & 0.51 & 104.6$\pm$12.0 & 355$\pm$2 & 206.5$\pm$23.6 & \\
29594AB & \nodata & 13.2 & 2013 Apr 22 & \nodata & 38.10 & 12834$\pm$1466 & 269 & 336.8$\pm$38.5 & L12\\
30193AB & \nodata & 14.6 & 2013 Apr 21 & 1.99 & 0.95 & 304.8$\pm$34.8 & 304$\pm$2 & 321.5$\pm$36.7  & \\
30838AB & 30837 & 12.5 & 2013 Apr 22 & 5.69 & 16.25 & 4436$\pm$507 & 25$\pm$2 & 273$\pm$31.2 & S02\\
31240AB & \nodata & 15.0 & 2013 Apr 21 & 4.16 & 0.74 & 251.2$\pm$28.7 & 210$\pm$2 & 338.3$\pm$38.7 &  \\
31240AC & \nodata & 15.0 & 2013 Apr 21 & 3.86 & 10.32 & 3491$\pm$399 & 157$\pm$2 & 338.3$\pm$38.7 & \\
34051AB & \nodata & 13.5 & 2013 Jan 19 & \nodata & \nodata & \nodata & \nodata & 242.3$\pm$27.7 & SB9\\
37342AB & 37341 & 14.4 & 2013 Apr 22 & 1.37 & 5.75 & 123.4$\pm$14.1 & 54$\pm$2 & 21.4$\pm$2.5  & S02\\
45616AB & \nodata & 11.9 & 2012 Sep 3 & 2.59 & 28.31 & 4696$\pm$536.8 & 113$\pm$2 & 165.9$\pm$19.0  & SB9\\
49486AB & 49487 & 15.9 & 2012 Oct 4 & 1.48 & 4.51 & 390.3$\pm$44.6 & 148$\pm$2 & 86.4$\pm$9.9  & S02\\
49819AB & 49821 & 14.0 & 2013 Aug 19 & 1.12 & 25.28 & 10263$\pm$1173 & 84$\pm$2 & 406$\pm$46.4 &  S02\\
50759AB & 50751 & 15.9 & 2012 Sep 13 & \nodata & 297.7 & 79156$\pm$9046 & 267.7 & 265.8$\pm$30.4 & S02\\
50869AB & \nodata & 15.8 & 2013 Aug 8 & 3.15 & 8.17 & 1707$\pm$195 & 19$\pm$2 & 209.0$\pm$24.0 & \\
52377AB & \nodata & 14.5 & 2012 Sep 4 & 2.35 & 0.92 & 561.3$\pm$64.2 & 211$\pm$2 & 585.3$\pm$66.9  & \\
52532AB & \nodata & 15.5 & 2012 Sep 4 & 2.60 & 0.30 & 52.82$\pm$6.0 & 168$\pm$2 & 175$\pm$20.0 & \\
52532AC & 52538 & 15.5 & 2012 Sep 4 & 3.35 & 37.14 & 6536$\pm$780 & \nodata & 176$\pm$21.0  & L79\\
53255AB & \nodata & 15.0 & 2013 Aug 16 & 0.75 & 1.07 & 123.9$\pm$14.2 & 68$\pm$2 & 112.7$\pm$12.9  & \\
53255AC & 53254 & 15.0 & 2013 Aug 16 & \nodata & 53.8 & 6063$\pm$694 & \nodata & 112.7$\pm$12.9  & L79\\
55603AB & \nodata & 12.1 & 2013 Aug 18 & 3.54 & 4.45 & 886.9$\pm$101.4 & 29$\pm$2 & 199.2$\pm$22.8  & \\
56818AB & \nodata & 14.0 & 2012 Sep 3 & 2.04 & 0.63 & 169.8$\pm$19.4 & 44$\pm$2 & 246.2$\pm$28.1 & \\
57038AB & \nodata & 13.9 & 2013 Aug 16 & 0.19 & 8.14 & 2508$\pm$286.7 & 335$\pm$2 & 308.3$\pm$35.2 & \\
57452AB & \nodata & 13.6 & 2013 Aug 16 & 1.91 & 1.98 & 474.5$\pm$54.2 & 77$\pm$2 & 234.9$\pm$26.9 & \\
57856AB & \nodata & 13.2 & 2013 Aug 17 & 5.08 & 2.00 & 585.3$\pm$66.9 & 169$\pm$2 & 289.7$\pm$33.1  & \\
58812AB & 58813 & 15.0 & 2013 Aug 16 & 1.40 & 2.81 & 743.6$\pm$85.0 & 69$\pm$2 & 264.4$\pm$30.2  & \\
\hline
\end{tabular}
\small
\label{tab:measurements}
\begin{flushleft}
Notes. --- References for previous detections are denoted using the following codes: Pourbaix et al. 2004 (SB9); Luyten 1979 (L79); Samir et al. 2002 (S02); L\'{o}pez et al. 2012 (L12).
\end{flushleft}
\end{table*}

\begin{figure}
\centering
\includegraphics[width=245pt]{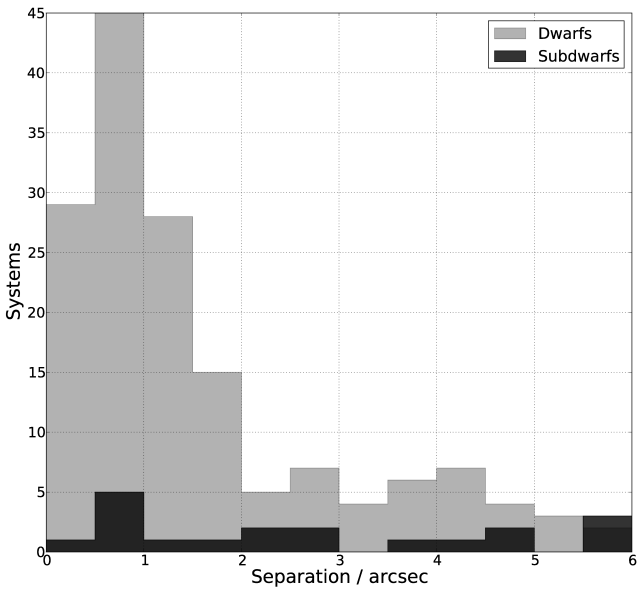}
\caption{Histogram of the angular separations of our subdwarf companions and the dwarf companions found by \citet{janson12}.  Only systems resolvable in both surveys were plotted (0$\farcs 15< \rho < 6 \farcs$0)}
\label{fig:hist_sep}
\end{figure}

The lack of close subdwarf companions has been noted previously by \citet{jao09} and by \citet{abt08}, however with significantly smaller samples.    A direct comparison of orbital separations is biased by the relative distance variation in the two samples.  With their relative rarity in the solar neighborhood, the subdwarf sample is overall approximately a factor of 4 further distant than the dwarf sample.  If the populations were similiar, this would result in a relative abundance of tight dwarf binaries, while the 6$\arcsec$ limit of the Janson et al. survey reduces the number of observed wide dwarf binaries.  Attempts to pick out similar systems by relative distance or by orbital separation from the two surveys results in a small statistical sample.  Nonetheless, the relative lack of close stars in the subdwarfs sample, as illustrated in Figure$~\ref{fig:hist_sep}$, and confirmed at high-confidence in our survey, warrants further investigation.

\subsection{Binarity and Metallicity}

The binary fraction we have found further confirms what has been suspected by past studies: that the binary fraction of subdwarfs is substantially lower than their dwarf cousins.  The largest survey of cool subdwarfs, although limited by the low angular resolution of the SDSS, \citet{zhang13}, find a multiplicity for type late K and M subdwarfs of 2.41\%, with an estimated lower bound of 10\% when adjusting for survey incompleteness.  This estimate and our work leave subdwarfs multiplicity rates approximately a factor of 2 to 4 lower than solar-metallicity stars of the same spectral types.

Historically, it has been a widely held view that metal-poor stars possess fewer stellar companions \citep{batten73,  latham04}.  A deficiency of eclipsing binaries was found in globular clusters by \citet{kopal59}, while \citet{jaschek59} discovered a deficiency of spectroscopic binaries in a sample of high-velocity dwarfs. \citet{abt87} used higher resolution CCD spectra to conclude that the frequency of spectroscopic binaries in high-velocity stars was half of metal-rich stars.  Recently, however, this view has come under attack.  \citet{carney94} used radial velocity measurements of 1464 stars, along with metallicity data \citep{carney87}, and found the difference in binary frequency of metal-rich and metal-poor stars to not be significant. Likewise, \citet{grether07} found a $\sim$2$\sigma$ anti-correlation between metallicity and companion stars.

In recent years, the relationship between planetary systems and metallicity has also been explored. \citet{fischer04} found a positive correlation between planetary systems and the metallicity of the host star.  This correlation has been reinforced to $\sim$4$\sigma$ by \citet{grether07}. Recently, \citet{wang14} found that planets in multiple-star systems occur 4.5$\pm$3.2, 2.6$\pm$1.0, and 1.7$\pm$0.5 times less frequently when the companion star is separated by 10, 100, and 1000 AU, respectively.

The solution may lie in the differences between halo and thick disk stars.  \citet{latham02} found no obvious difference between the binary fraction of the two populations, however \citet{chiba00} found a 55\% multiplicity rate for thick disk stars and 12\% for halo stars. \citeauthor{grether07} also find that the thick disk shows a $\sim$4 times higher binary fraction than halo stars, further hypothesizing that the mixing of the populations is the explanation for the perceived anti-correlation of metallicity and binarity.

The large difference between the M subdwarfs and thick-disk M dwarfs, apparent in our work in this paper and \citet{janson12}, seems to imply the two populations formed under different initial conditions.  Star formation in less dense regions appear to lower binary rates.  \citet{kohler06} found a factor 3-5 difference in binary fraction between the low-density Taurus star-forming region and the dense Orion cluster. It is also possible that, as older than solar-abundance stars, the metal-poor subdwarfs could have suffered more disruptive encounters with other stars.  These disturbances could separate companions with separations larger than a few AU, with the tighter, more highly bound systems being less affected \citep{sterzik98, abt08}, a theory derived from $N$-body simulations \citep{aarseth72, kroupa95}.  This, however, is contrary to our tentative result of a lack of close subdwarf companions, and the similar observations of \citet{jao09} and \citet{abt08} that close subdwarf binaries are rare. This implies that metal-poor subdwarfs had shorter lifetimes in clusters than their younger, metal-rich cousins, either being ejected or formed in a disrupted cluster.

Another possible explanation is that a large number of low-metallicity stars in the Milky Way could have resulted from past mergers with satellite galaxies.  Simulations from \citet{abadi06} predict that the early Galaxy underwent a period of active merging.  From these mergers, the Galaxy would inherit large numbers of metal-poor stars.  \citet{meza05} observe a group of metal-poor stars with angular momenta similar to the cluster $\omega$ Cen, long theorized to be the core of a dwarf galaxy that merged with the Milky Way.  The environment of these foreign galaxies is unknown, so star formation could be quite different than our own Galaxy.  It is also possible that during the merger multiple close stellar encounters and perturbations could alter their primordial binary properties.

\section{Conclusions}
\label{sec:Conclusions}

In the largest high-resolution binary survey of cool subdwarfs, we observed 344 stars with the Robo-AO robotic laser adaptive optics system, sensitive to companions at $\rho \ge 0\farcs$15 and $\Delta m_i \le$ 6.  Of those targets, we observed 16 new multiple systems and 4 new companions to already known binary systems.  When including previously recorded multiple systems, this implies a multiplicity rate for cool subdwarfs of $11.6\%\pm1.8\%$ and a triplet fraction of 1.7$\%\pm .7 \%$.  This is significantly lower than the observed cool subdwarf binarity of 26$\%\pm 6 \%$ by \citet{jao09} and in agreement with the completeness adjusted estimate of $> 10 \%$ of \citet{zhang13}. When comparing our results to similar surveys of dwarf binarity, we note a $\sim$2.8$\sigma$ difference in relative magnitude differences between companions.  An apparent lack of close binaries is noted, as has been previously observed in the literature.  The high efficiency of Robo-AO makes large, high-angular resolution surveys practical and will in the future continue to put tighter constraints on the properties of stellar populations.

\section*{Acknowledgements}
The Robo-AO system is supported by collaborating partner institutions, the California Institute of Technology and the Inter-University Centre for Astronomy and Astrophysics, and by the National Science Foundation under Grant Nos. AST-0906060, AST-0960343, and AST-1207891, by the Mount Cuba Astronomical Foundation, by a gift from Samuel Oschin. 

We are grateful to the Palomar Observatory staff for their ongoing support of Robo-AO on the 60 inch telescope, particularly S. Kunsman, M. Doyle, J. Henning, R. Walters, G. Van Idsinga, B. Baker, K. Dunscombe and D. Roderick.  The SOAR telescope is operated by the Association of Universities for Research in Astronomy, Inc., under a cooperative agreement between the CNPq, Brazil, the National Observatory for Optical Astronomy (NOAO), the University of North Carolina, and Michigan State University, USA.  We also thank the SOAR operators, notably  Sergio Pizarro. We recognize and acknowledge the very significant cultural role and reverence that the summit of Maunakea has always had within the indigenous Hawaiian community. We are most fortunate to have the opportunity to conduct observations from this mountain.  

C.B. acknowledges support from the Alfred P. Sloan Foundation. 

This research has made use of the SIMBAD database, operated by Centre des Donn\'ees Stellaires (Strasbourg, France), and bibliographic references from the Astrophysics Data System maintained by SAO/NASA.

{\it Facilities:} \facility{PO:1.5m (Robo-AO)}, \facility{Keck:II (NIRC2-LGS)}, \facility{SOAR (Goodman)}

\section{appendix}

In Table$~\ref{tab:roboaolist}$, we list our Robo-AO observed subdwarfs, including date the target was observed, observation quality as described in Section Section \ref{sec:imageperf}, and the presence of detected companions.

\bigskip
\bigskip
\bigskip

\begin{center}
\begin{longtable}{cclcc}
\caption{Full Robo-AO Observation List}\\
\hline
\hline
\noalign{\vskip 3pt}  
\text{NLTT} & \textit{m$_v$} & \text{ObsID} & \text{Obs. qual}  & \text{Companion?}\\ [0.3ex]
\hline
\noalign{\vskip 3pt}  
\endfirsthead
\multicolumn{5}{c}
{\tablename\ \thetable\ -- \textit{Continued}} \\
\hline
\noalign{\vskip 3pt} 
\text{NLTT} & \textit{m$_v$} & \text{ObsID} & \text{Obs. qual}  & \text{Companion?}\\ [0.3ex]
\hline
\noalign{\vskip 3pt}  
\endhead
\endfoot
\hline
\endlastfoot
69 & 15.2 & 2012 Oct 10 & low & \\ 
193 & 15.5 & 2013 Aug 15 & medium & \\ 
341 & 12.1 & 2012 Oct 10 & high & \\ 
361 & 15.4 & 2013 Aug 17 & low & \\ 
496 & 15.8 & 2012 Sep 04 & medium & \\ 
660 & 15.7 & 2012 Sep 03 & low & \\ 
812 & 12.8 & 2012 Sep 03 & high & \\ 
933 & 15.5 & 2013 Aug 16 & low & \\ 
1020 & 15.3 & 2013 Aug 15 & medium & \\ 
1059 & 13.8 & 2012 Sep 04 & medium & \\ 
1231 & 11.9 & 2013 Aug 16 & high & \\ 
1509 & 15.8 & 2013 Aug 16 & low & \\ 
1575 & 16.2 & 2012 Sep 03 & low & \\ 
1635 & 13.2 & 2012 Sep 03 & high & \\ 
1684 & 15.1 & 2012 Sep 13 & low & \\ 
1815 & 15.5 & 2012 Sep 04 & low & \\ 
1870 & 13.9 & 2012 Sep 03 & medium & \\ 
2045 & 13.5 & 2013 Aug 15 & medium & yes\\ 
2107 & 15.5 & 2012 Sep 04 & low & \\ 
2205 & 14.0 & 2013 Aug 15 & medium & yes\\ 
2324 & 15.7 & 2013 Aug 16 & medium & yes\\ 
2868 & 13.5 & 2013 Aug 16 & medium & \\ 
2953 & 15.9 & 2012 Sep 04 & low & \\ 
2966 & 15.6 & 2012 Sep 04 & medium & \\ 
3035 & 15.9 & 2012 Sep 04 & low & \\ 
3965 & 16.1 & 2013 Aug 16 & medium & \\ 
4245 & 15.6 & 2013 Aug 15 & low & \\ 
4447 & 15.9 & 2012 Sep 03 & low & \\ 
4817 & 11.4 & 2012 Sep 03 & high & yes\\ 
4838 & 15.4 & 2012 Sep 03 & low & \\ 
5022 & 13.9 & 2012 Sep 03 & medium & \\ 
5192 & 14.3 & 2012 Sep 03 & medium & \\ 
5289 & 15.6 & 2012 Sep 03 & low & \\ 
6519 & 14.8 & 2012 Sep 03 & medium & \\ 
6582 & 15.7 & 2013 Aug 17 & low & \\ 
6614 & 15.7 & 2012 Sep 03 & medium & \\ 
6816 & 16.1 & 2013 Aug 15 & low & \\ 
6856 & 16.1 & 2012 Sep 03 & low & \\ 
6863 & 15.3 & 2013 Aug 17 & low & \\ 
7078 & 14.4 & 2012 Sep 03 & medium & \\ 
7207 & 14.5 & 2013 Aug 15 & medium & \\ 
7299 & 11.5 & 2013 Aug 16 & high & \\ 
7301 & 14.9 & 2012 Sep 03 & high & yes\\ 
7415 & 9.1 & 2012 Sep 03 & high & \\ 
7417 & 11.6 & 2013 Aug 15 & high & \\ 
7467 & 15.9 & 2012 Sep 13 & low & \\ 
7596 & 16.2 & 2013 Aug 17 & low & \\ 
7654 & 16.1 & 2013 Aug 16 & medium & \\ 
7769 & 14.0 & 2012 Sep 03 & medium & \\ 
7914 & 14.3 & 2012 Sep 03 & medium & yes\\ 
8034 & 11.8 & 2012 Sep 03 & high & \\ 
8227 & 10.5 & 2013 Aug 17 & high & \\ 
8342 & 14.9 & 2012 Sep 03 & medium & \\ 
8405 & 15.8 & 2012 Sep 03 & medium & \\ 
8507 & 13.9 & 2012 Sep 03 & medium & \\ 
8783 & 11.5 & 2012 Sep 03 & high & \\ 
8866 & 15.8 & 2013 Aug 16 & low & \\ 
9523 & 15.4 & 2013 Aug 15 & low & \\ 
9550 & 15.5 & 2013 Aug 19 & low & \\ 
9578 & 10.5 & 2013 Aug 15 & high & \\ 
9597 & 12.0 & 2012 Sep 13 & high & \\ 
9622 & 14.3 & 2012 Sep 04 & medium & \\ 
9648 & 14.9 & 2012 Sep 04 & medium & \\ 
9653 & 15.6 & 2013 Aug 16 & low & \\ 
9727 & 15.8 & 2013 Aug 15 & medium & \\ 
9734 & 15.0 & 2012 Sep 04 & medium & \\ 
9799 & 15.4 & 2012 Sep 13 & low & \\ 
9848 & 16.6 & 2013 Aug 19 & low & \\ 
9898 & 14.2 & 2013 Aug 19 & low & \\ 
9938 & 16.2 & 2013 Aug 15 & low & \\ 
10018 & 15.4 & 2013 Aug 17 & low & \\ 
10022 & 15.8 & 2013 Aug 16 & medium & \\ 
10135 & 15.7 & 2012 Sep 04 & low & \\ 
10176 & 15.8 & 2013 Aug 20 & low & \\ 
10243 & 14.1 & 2012 Sep 04 & medium & \\ 
10401 & 14.6 & 2013 Aug 18 & low & \\ 
10517 & 14.5 & 2012 Sep 04 & medium & \\ 
10536 & 11.2 & 2013 Aug 15 & high & yes\\ 
10548 & 15.9 & 2013 Aug 15 & low & \\ 
10850 & 10.7 & 2012 Sep 04 & high & \\ 
10883 & 15.9 & 2012 Sep 04 & low & \\ 
11007 & 12.2 & 2013 Aug 21 & high & \\ 
11010 & 14.1 & 2012 Sep 04 & medium & \\ 
11015 & 16.3 & 2013 Aug 16 & low & yes\\ 
11032 & 14.2 & 2012 Sep 04 & medium & \\ 
11068 & 15.4 & 2013 Aug 21 & low & \\ 
11938 & 14.3 & 2012 Sep 04 & medium & \\ 
12017 & 12.3 & 2013 Aug 17 & high & \\ 
12026 & 15.8 & 2013 Aug 18 & low & \\ 
12044 & 15.8 & 2012 Sep 13 & low & \\ 
12227 & 14.2 & 2013 Aug 18 & medium & \\ 
12350 & 12.1 & 2013 Aug 18 & medium & \\ 
12489 & 14.6 & 2012 Oct 10 & low & \\ 
12537 & 14.5 & 2013 Aug 21 & medium & \\ 
12704 & 15.4 & 2012 Oct 10 & low & \\ 
12769 & 14.1 & 2013 Aug 18 & medium & \\ 
12829 & 14.6 & 2012 Oct 03 & medium & \\ 
12845 & 10.6 & 2012 Oct 03 & high & yes\\ 
12856 & 10.8 & 2013 Aug 18 & high & \\ 
12876 & 15.6 & 2012 Oct 03 & low & \\ 
12923 & 15.2 & 2013 Aug 18 & low & \\ 
13022 & 15.9 & 2012 Oct 03 & low & \\ 
13344 & 13.8 & 2012 Oct 03 & medium & \\ 
13368 & 15.5 & 2012 Oct 03 & low & \\ 
13402 & 14.7 & 2012 Oct 03 & low & \\ 
13469 & 15.1 & 2013 Aug 18 & low & \\ 
13470 & 13.8 & 2012 Oct 03 & medium & \\ 
13641 & 12.9 & 2012 Oct 06 & high & \\ 
13660 & 12.4 & 2012 Oct 03 & high & \\ 
13694 & 15.4 & 2013 Aug 20 & medium & \\ 
13706 & 14.5 & 2012 Oct 03 & low & \\ 
13770 & 12.4 & 2012 Oct 03 & high & \\ 
13811 & 13.4 & 2012 Oct 03 & medium & \\ 
13920 & 14.4 & 2013 Aug 20 & medium & \\ 
13940 & 14.4 & 2012 Oct 05 & medium & \\ 
14091 & 13.9 & 2012 Oct 05 & medium & \\ 
14131 & 13.4 & 2012 Oct 03 & medium & \\ 
14169 & 13.4 & 2012 Oct 05 & medium & \\ 
14197 & 12.4 & 2012 Oct 04 & low & \\ 
14391 & 13.5 & 2012 Oct 04 & low & \\ 
14450 & 14.7 & 2012 Oct 04 & low & \\ 
14549 & 14.5 & 2012 Oct 10 & low & \\ 
14822 & 12.7 & 2012 Oct 03 & medium & \\ 
14864 & 14.3 & 2012 Oct 07 & low & \\ 
15039 & 14.8 & 2012 Oct 10 & low & \\ 
15183 & 12.6 & 2012 Oct 07 & medium & \\ 
15218 & 12.3 & 2012 Oct 06 & high & \\ 
15973 & 9.3 & 2012 Oct 07 & high & yes\\ 
15974 & 13.8 & 2012 Oct 07 & high & \\ 
16030 & 13.9 & 2012 Oct 07 & low & \\ 
16185 & 14.4 & 2012 Oct 10 & low & \\ 
16242 & 10.6 & 2012 Oct 06 & medium & \\ 
16579 & 12.3 & 2012 Oct 09 & high & \\ 
16606 & 12.3 & 2012 Oct 10 & high & \\ 
16849 & 15.3 & 2012 Oct 10 & low & \\ 
16869 & 13.2 & 2013 Jan 20 & high & \\ 
16986 & 15.8 & 2013 Jan 20 & low & \\ 
17039 & 12.9 & 2012 Oct 10 & medium & \\ 
17485 & 11.9 & 2012 Oct 10 & high & yes\\ 
17680 & 13.6 & 2013 Jan 20 & medium & \\ 
17786 & 12.0 & 2013 Jan 20 & high & \\ 
17872 & 10.7 & 2013 Jan 20 & high & \\ 
18019 & 13.3 & 2012 Oct 10 & medium & \\ 
18131 & 14.4 & 2013 Jan 20 & medium & \\ 
18424 & 12.7 & 2013 Jan 18 & high & \\ 
18463 & 13.8 & 2013 Jan 20 & high & \\ 
18502 & 12.2 & 2013 Jan 19 & high & yes\\ 
18731 & 13.1 & 2013 Jan 19 & high & \\ 
18798 & 14.5 & 2013 Jan 19 & high & yes\\ 
18799 & 11.0 & 2013 Jan 19 & high & \\ 
19037 & 14.9 & 2013 Jan 20 & medium & \\ 
19210 & 11.2 & 2013 Jan 20 & high & yes\\ 
19301 & 14.7 & 2013 Jan 19 & low & \\ 
19570 & 14.4 & 2013 Apr 22 & medium & \\ 
19614 & 15.7 & 2013 Apr 22 & medium & \\ 
19643 & 11.9 & 2013 Jan 19 & high & \\ 
19824 & 14.6 & 2013 Jan 19 & medium & \\ 
20252 & 14.9 & 2013 Apr 22 & medium & \\ 
20288 & 14.9 & 2013 Apr 22 & medium & \\ 
20392 & 13.8 & 2013 Jan 22 & low & \\ 
20476 & 13.2 & 2013 Apr 22 & high & \\ 
20492 & 13.3 & 2013 Jan 19 & high & \\ 
20684 & 12.0 & 2013 Jan 19 & high & \\ 
20691 & 9.6 & 2013 Jan 19 & high & yes\\ 
20768 & 14.0 & 2013 Jan 19 & medium & \\ 
21039 & 14.0 & 2013 Jan 19 & medium & \\ 
21112 & 15.3 & 2013 Apr 22 & medium & \\ 
21133 & 12.7 & 2013 Jan 19 & medium & \\ 
21341 & 14.3 & 2013 Jan 19 & low & \\ 
21370 & 13.7 & 2013 Jan 19 & medium & yes\\ 
21449 & 12.6 & 2013 Apr 22 & high & \\ 
21601 & 14.6 & 2013 Apr 22 & medium & \\ 
22026 & 12.6 & 2013 Apr 22 & high & \\ 
22053 & 12.1 & 2013 Jan 19 & high & \\ 
22520 & 10.8 & 2013 Jan 19 & high & \\ 
22752 & 13.9 & 2013 Jan 19 & medium & \\ 
22945 & 13.2 & 2013 Apr 22 & medium & \\ 
23894 & 14.6 & 2013 Jan 18 & low & \\ 
24006 & 15.5 & 2013 Apr 22 & medium & \\ 
24082 & 13.1 & 2013 Jan 19 & medium & yes\\ 
24353 & 13.2 & 2013 Jan 18 & medium & \\ 
24371 & 14.2 & 2013 Jan 18 & low & \\ 
24718 & 13.1 & 2013 Jan 18 & medium & \\ 
24984 & 12.5 & 2013 Apr 21 & high & \\ 
25006 & 14.1 & 2013 Apr 21 & medium & \\ 
25177 & 12.2 & 2013 Apr 22 & high & \\ 
25190 & 13.9 & 2013 Jan 18 & low & \\ 
25234 & 13.2 & 2013 Jan 18 & medium & yes\\ 
25475 & 13.9 & 2013 Apr 21 & medium & \\ 
25776 & 13.8 & 2013 Apr 22 & medium & \\ 
25909 & 13.5 & 2013 Apr 22 & high & \\ 
25970 & 14.9 & 2013 Jan 18 & low & \\ 
26232 & 14.4 & 2013 Jan 18 & low & \\ 
26482 & 12.5 & 2013 Jan 18 & medium & \\ 
26503 & 14.2 & 2013 Apr 21 & medium & \\ 
26532 & 14.8 & 2013 Jan 18 & low & \\ 
26565 & 14.8 & 2013 Jan 18 & low & \\ 
26588 & 13.6 & 2013 Apr 21 & high & \\ 
26677 & 13.5 & 2013 Jan 18 & low & \\ 
27436 & 13.0 & 2013 Jan 18 & medium & \\ 
27763 & 13.6 & 2013 Jan 18 & medium & \\ 
27767 & 14.7 & 2013 Apr 21 & medium & \\ 
28199 & 13.2 & 2013 Jan 18 & medium & \\ 
28304 & 13.3 & 2013 Apr 22 & medium & \\ 
28434 & 14.9 & 2013 Jan 17 & low & yes\\ 
29023 & 13.0 & 2013 Jan 18 & medium & \\ 
29064 & 14.0 & 2013 Apr 21 & medium & \\ 
29256 & 14.7 & 2013 Jan 18 & low & \\ 
29442 & 14.4 & 2013 Jan 18 & low & \\ 
29551 & 11.5 & 2013 Apr 21 & high & yes\\ 
29594 & 13.2 & 2013 Apr 22 & high & yes\\ 
29933 & 10.2 & 2013 Apr 22 & high & \\ 
30128 & 13.1 & 2013 Apr 21 & high & \\ 
30193 & 14.6 & 2013 Apr 21 & medium & yes\\ 
30462 & 12.8 & 2013 Jan 18 & medium & \\ 
30636 & 14.8 & 2013 Jan 18 & low & \\ 
30824 & 14.6 & 2013 Jan 17 & low & \\ 
30838 & 12.5 & 2013 Apr 22 & high & yes\\ 
31146 & 12.0 & 2013 Apr 21 & high & \\ 
31155 & 13.6 & 2013 Jan 18 & medium & \\ 
31240 & 15.0 & 2013 Apr 21 & medium & yes\\ 
31965 & 14.2 & 2013 Jan 19 & medium & \\ 
32316 & 11.3 & 2013 Apr 22 & high & \\ 
32392 & 14.6 & 2013 Jan 19 & medium & \\ 
32562 & 14.3 & 2013 Jan 17 & low & \\ 
32648 & 12.8 & 2013 Jan 18 & medium & \\ 
32917 & 13.8 & 2013 Apr 22 & medium & \\ 
32995 & 13.4 & 2013 Apr 22 & high & \\ 
33104 & 14.0 & 2013 Jan 18 & low & \\ 
33156 & 14.2 & 2013 Apr 22 & medium & \\ 
33371 & 12.8 & 2013 Jan 17 & medium & \\ 
33971 & 12.8 & 2013 Jan 18 & medium & \\ 
34051 & 13.5 & 2013 Jan 19 & low & yes\\ 
34628 & 11.9 & 2013 Apr 21 & high & \\ 
35068 & 13.2 & 2013 Jan 18 & medium & \\ 
35318 & 13.4 & 2013 Apr 21 & high & \\ 
36020 & 14.2 & 2013 Apr 22 & medium & \\ 
37342 & 14.4 & 2013 Apr 22 & high & yes\\ 
37684 & 13.3 & 2013 Apr 22 & high & \\ 
37807 & 12.0 & 2013 Apr 22 & high & \\ 
39378 & 13.5 & 2013 Apr 22 & high & \\ 
39721 & 13.6 & 2013 Apr 22 & high & \\ 
40022 & 13.9 & 2013 Apr 22 & medium & \\ 
40313 & 13.7 & 2013 Apr 22 & high & \\ 
41111 & 13.7 & 2013 Apr 22 & medium & \\ 
44039 & 11.5 & 2012 Sep 14 & high & \\ 
44233 & 15.2 & 2012 Sep 04 & low & \\ 
44568 & 12.3 & 2012 Sep 04 & high & \\ 
44639 & 11.8 & 2012 Sep 04 & high & \\ 
44769 & 15.2 & 2013 Apr 21 & medium & \\ 
45609 & 12.5 & 2012 Sep 04 & high & \\ 
45616 & 11.9 & 2012 Sep 04 & high & yes\\ 
47480 & 13.8 & 2012 Oct 05 & low & \\ 
47543 & 9.2 & 2012 Oct 05 & medium & \\ 
48011 & 14.7 & 2012 Oct 05 & high & \\ 
48056 & 13.7 & 2012 Oct 07 & low & \\ 
48391 & 15.2 & 2012 Oct 05 & medium & \\ 
48592 & 12.2 & 2012 Oct 04 & medium & \\ 
48866 & 12.7 & 2012 Oct 04 & medium & \\ 
49486 & 16.0 & 2012 Oct 04 & medium & yes\\ 
49487 & 12.3 & 2012 Oct 04 & medium & \\ 
49488 & 14.9 & 2013 Aug 19 & medium & \\ 
49618 & 12.2 & 2012 Oct 04 & medium & \\ 
49726 & 15.9 & 2013 Aug 19 & low & \\ 
49749 & 14.8 & 2012 Oct 03 & medium & \\ 
49819 & 14.0 & 2013 Aug 19 & high & yes\\ 
49821 & 12.8 & 2013 Aug 19 & high & \\ 
49897 & 15.8 & 2012 Oct 04 & low & \\ 
50257 & 13.8 & 2013 Aug 18 & low & \\ 
50376 & 13.9 & 2012 Sep 13 & medium & \\ 
50556 & 15.7 & 2012 Sep 13 & low & \\ 
50759 & 15.9 & 2012 Sep 13 & low & yes\\ 
50869 & 15.8 & 2013 Aug 19 & low & \\ 
50911 & 11.6 & 2012 Sep 13 & high & \\ 
51006 & 14.1 & 2013 Aug 19 & medium & \\ 
51153 & 15.1 & 2012 Sep 13 & low & \\ 
51740 & 15.3 & 2012 Sep 13 & low & \\ 
51754 & 15.0 & 2012 Sep 13 & low & \\ 
51824 & 11.9 & 2013 Aug 18 & medium & \\ 
51856 & 13.4 & 2012 Sep 04 & medium & \\ 
52089 & 14.9 & 2012 Sep 04 & medium & \\ 
52377 & 14.5 & 2012 Sep 04 & medium & yes\\ 
52532 & 15.5 & 2012 Sep 04 & low & yes\\ 
52573 & 15.3 & 2013 Aug 18 & low & \\ 
52666 & 15.0 & 2013 Aug 19 & low & \\ 
52816 & 15.7 & 2012 Sep 13 & low & \\ 
52894 & 16.0 & 2012 Sep 13 & low & \\ 
53190 & 15.4 & 2013 Aug 16 & medium & \\ 
53254 & 14.7 & 2013 Aug 16 & medium & \\ 
53255 & 15.0 & 2013 Aug 16 & medium & yes\\ 
53274 & 11.9 & 2013 Aug 17 & high & \\ 
53316 & 15.4 & 2012 Sep 13 & low & \\ 
53346 & 13.8 & 2013 Aug 17 & medium & \\ 
53480 & 12.6 & 2013 Aug 17 & high & \\ 
53702 & 15.3 & 2012 Sep 13 & medium & \\ 
53707 & 12.1 & 2013 Aug 18 & medium & \\ 
53781 & 13.8 & 2013 Aug 17 & medium & \\ 
53801 & 11.8 & 2012 Sep 13 & high & \\ 
53823 & 13.8 & 2013 Aug 18 & low & \\ 
54027 & 13.3 & 2013 Aug 19 & medium & \\ 
54088 & 14.1 & 2013 Aug 18 & low & \\ 
54168 & 13.4 & 2013 Aug 17 & medium & \\ 
54184 & 14.0 & 2013 Aug 17 & medium & \\ 
54349 & 14.4 & 2012 Sep 13 & medium & \\ 
54450 & 15.6 & 2013 Aug 16 & low & \\ 
54578 & 15.8 & 2013 Aug 18 & low & \\ 
54608 & 16.0 & 2013 Aug 16 & low & \\ 
54620 & 15.2 & 2013 Aug 17 & medium & \\ 
54699 & 15.1 & 2012 Sep 13 & low & \\ 
54710 & 15.2 & 2012 Sep 13 & low & \\ 
54730 & 11.5 & 2012 Sep 13 & high & \\ 
55411 & 15.9 & 2013 Aug 16 & low & \\ 
55603 & 12.1 & 2013 Aug 18 & medium & yes\\ 
55732 & 13.4 & 2013 Aug 17 & medium & \\ 
55733 & 14.5 & 2012 Sep 03 & medium & \\ 
55942 & 13.5 & 2013 Aug 16 & medium & \\ 
56002 & 14.4 & 2012 Sep 03 & medium & \\ 
56290 & 12.6 & 2013 Aug 16 & high & \\ 
56420 & 15.6 & 2012 Sep 03 & low & \\ 
56533 & 15.9 & 2013 Aug 16 & low & \\ 
56534 & 12.7 & 2013 Aug 17 & high & \\ 
56774 & 12.9 & 2013 Aug 18 & low & \\ 
56817 & 16.1 & 2013 Aug 17 & low & \\ 
56818 & 14.0 & 2012 Sep 03 & medium & yes\\ 
56855 & 13.7 & 2013 Aug 16 & medium & \\ 
57038 & 13.9 & 2013 Aug 16 & medium & yes\\ 
57214 & 15.8 & 2013 Aug 16 & low & \\ 
57452 & 13.6 & 2013 Aug 16 & medium & yes\\ 
57546 & 16.2 & 2013 Aug 17 & low & \\ 
57564 & 10.6 & 2013 Aug 17 & high & \\ 
57630 & 15.0 & 2013 Aug 16 & medium & \\ 
57631 & 13.5 & 2013 Aug 17 & medium & \\ 
57647 & 14.7 & 2013 Aug 17 & medium & \\ 
57741 & 14.2 & 2013 Aug 17 & medium & \\ 
57744 & 16.1 & 2013 Aug 17 & low & \\ 
57781 & 10.1 & 2013 Aug 16 & high & \\ 
57832 & 15.2 & 2012 Sep 03 & medium & \\ 
57851 & 15.2 & 2012 Sep 03 & medium & \\ 
57856 & 13.2 & 2013 Aug 17 & medium & yes\\ 
58071 & 13.1 & 2012 Sep 03 & medium & \\ 
58141 & 15.8 & 2013 Aug 16 & low & \\ 
58403 & 15.2 & 2013 Aug 16 & low & \\ 
58522 & 15.0 & 2013 Aug 17 & medium & \\ 
58555 & 15.1 & 2012 Sep 03 & medium & \\ 
58812 & 14.9 & 2013 Aug 16 & medium & yes
\label{tab:roboaolist}
\end{longtable}
\end{center}

\end{document}